\documentclass[aps,twocolumn]{revtex4}

\usepackage{graphicx}

\begin{document}

\title{Direct measurement and reconstruction of nonclassical features of
twin beams generated in spontaneous parametric down-conversion}

\author{Ond\v{r}ej Haderka$ $\thanks{e-mail:
haderka@sloup.upol.cz},
Jan Pe\v{r}ina Jr.,
Martin Hamar,
Jan Pe\v{r}ina, \\
Joint Laboratory of Optics
of Palack\'{y} University and \\
Institute of Physics of Academy of Sciences of the Czech
Republic, \\
17.~listopadu 50A, 772 00 Olomouc,
Czech Republic \\
}

\begin{abstract}
Correlations in twin beams composed of many photon pairs are
studied using an intensified CCD camera. Joint signal-idler
photon-number distribution and quantum phase-space quasi-distributions
determined from experimental data have nonclassical features.
\end{abstract}


\maketitle

Quantum mechanics interprets nonlinear optical processes as
physical effects composed of many elementary `quantum' events in
which one (several) photon is annihilated and several (one)
photon emerge \cite{Milburn1995}. This behavior of nonlinearly
interacting physical systems is completely unusual in classical
physics and lies in heart of nonclassical-light generation.
Generation of correlated photon pairs in the process of
spontaneous parametric frequency down-conversion is probably the
most common example \cite{Mandel1995}. Fundamental experiments
performed for the first time with photon pairs have shown that
these nonlinear elementary `quantum' events have even an `internal
structure'. Measurement using photon pairs in Hong-Ou-Mandel
interferometer has revealed this structure demonstrating that
photons in one photon pair (photon twins) are generated within a
sharp time window typically of several tens of fs \cite{Hong1987}.
Later experiments have even shown that entangled photon pairs are
the same fundamental entities as single photons.
Similarly as a single photon can interfere only with itself, an
entangled photon pair interferes only with itself \cite{Zou1991}.
These completely unusual properties of photon pairs have been
experimentally tested in numerous experiments with the same
qualitative conclusion - fundamental laws of quantum mechanics are
a solid basis for the explanation of obtained experimental
results. Among others, experimental confirmation of violation of
Bell inequalities using photon pairs has excluded neoclassical
theories with local hidden variables as a right tool for the description
of Nature. Photon pairs are also an indispensable tool in quantum
teleportation \cite{Bouwmeester1997}, quantum cryptography or
dense coding.

The use of intense femtosecond pump fields together with
availability of new materials with larger nonlinearities have
opened a new area in investigation of twin beams. Nowadays even
beams containing many photon pairs generated in a sharp time
window can be obtained. The first experiments confirm in agreement
with quantum theory that photon pairs inside such beams behave as
independent entities. Coherence properties of twin beams
reflect those of the pump beam. Moreover, it has been shown that
photon pairs can have their origin also in stimulated emission
\cite{Lamas2001} (laser-like generation of twin beams is
sometimes mentioned).

These fundamental properties of photon pairs then determine
statistical properties of `more intense' twin beams. In this
letter, we report on experimental determination of photocount
statistics of twin beams \cite{Larchuk1995}
using an intensified CCD camera (iCCD).
An alternative approach to determine photocount statistics
is based on homodyne detection and has revealed correlations
in photon numbers of two fields comprising a two-mode squeezed state
\cite{Vasilyev2000}.
In our case, even raw experimental data show nonclassical properties
of twin beams. The
reconstructed photon-number distribution then enables to determine
(radially symmetric) joint phase-space quasi-distributions
of the twin beams.

\begin{figure}         
 \centerline{\includegraphics[width=0.40\textwidth]{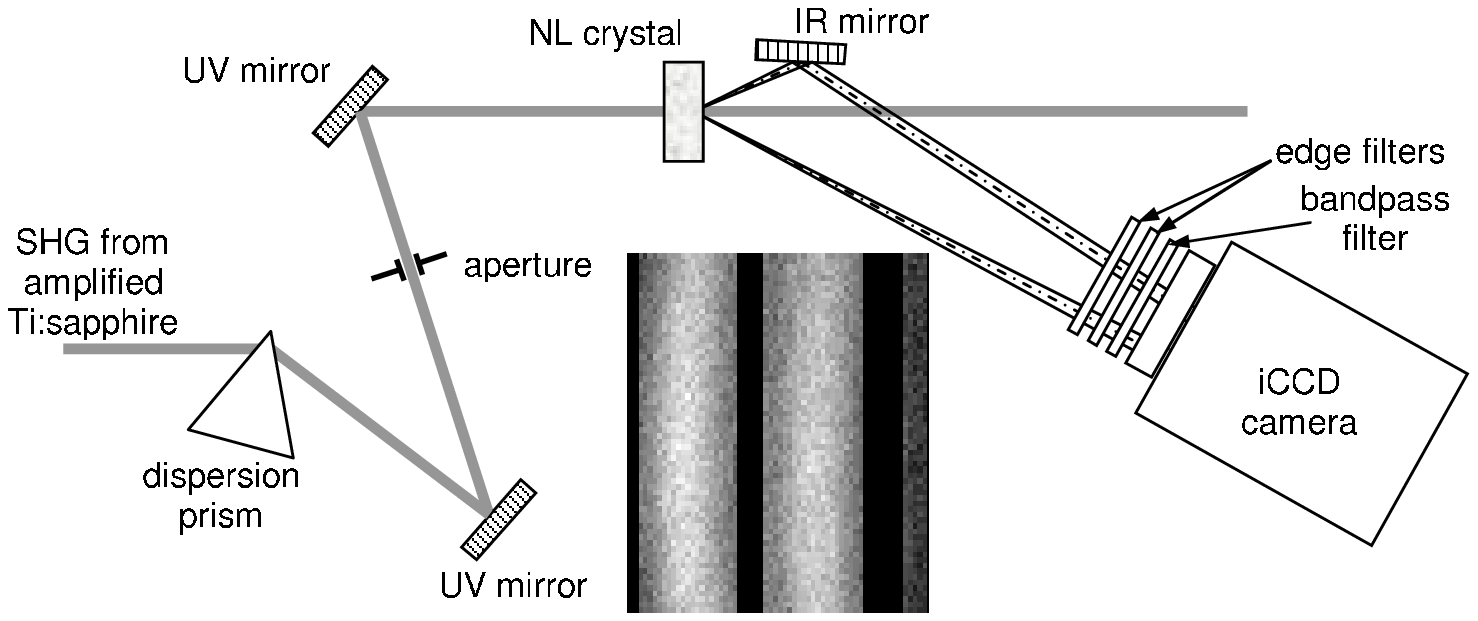}}
 \vspace{3mm}
 \caption{Scheme of the experiment. The inset shows accumulated
  picture after 240,000 frames.}
\label{setup}
\end{figure}

In the reported experiment sketched in Fig.~\ref{setup}, we use
the second-harmonic of an amplified femtosecond Ti:sapphire laser
system as pump source for the generation of down-converted light
in a 5-mm long Li:IO$_3$ nonlinear crystal.
Laser system operates at 800~nm and yields a train of 200~fs
pulses with a repetition rate controlled by the amplifier and set
to 11~kHz in our case. Their second-harmonic is produced in a 2-mm
long BBO crystal and pulses of energy up to 1.3~$\mu$J are
obtained. Down-converted photons emerge from a crystal at a cone
layer with vertex half-angle of 31~deg (see Fig.~\ref{setup}). One
part (signal) of the cone layer is captured by the photocathode of
the camera placed 10~cm behind the crystal. The corresponding
opposite part of the cone layer (idler) is led to the camera after
reflection on a high-reflectivity mirror placed as close to the
crystal as possible to minimize path difference of the two beams.
Prior to entering the camera, both beams are filtered by two edge
filters (high-pass above 750~nm) and a 20~nm wide (FWHM) bandpass
filter centered at 800~nm. The filters reduce the noise coming
from fluorescence in the nonlinear crystal and stray light in the
laboratory to an acceptable level and bandpass filter defines the
thickness of the cone layer. In the software of the camera, three
regions of interest are defined, two for the signal and idler
strips and a third one that serves for monitoring the noise level
(see inset in Fig.~\ref{setup}). The overall quantum efficiency of
the iCCD camera has been found to be 9.5\%.

Quantum state of twin beams at the output plane of the crystal
can be described by the following statistical operator
$\hat{\rho}_{SI}$ \cite{PerinaJr2003}:
\begin{equation}       
 \hat{\rho}_{SI} = \sum_{n_S=0}^{\infty} \sum_{n_I=0}^{\infty}
  p(n_S,n_I) |n_S\rangle_S {}_S \langle n_S| \otimes
  |n_I\rangle_I {}_I \langle n_I| ;
\end{equation}
$ |n_S\rangle_S  $ ($ |n_I\rangle_I $) denotes Fock state
with $ n_S $ ($ n_I $) photons and then $ p(n_S,n_I) $ means the
joint signal-idler photon-number distribution. The down-converted
photon pairs then undergo several loss mechanisms before they are
registered. Taking into account losses in the signal and idler
paths (expressed using effective transmittances $ T_S $, $ T_I $),
quantum efficiency of the camera ($ \eta $), and noise due to
other light sources and internal noises of the camera ($ D $), the
measured probabilities $ f(c_S,c_I) $ of having $ c_S $ detections
in the signal strip and $ c_I $ detections in the idler strip at
the camera are determined as:
\begin{eqnarray}    
 f(c_S,c_I) &=& \sum_{n_S=0}^{\infty} \sum_{n_I=0}^{\infty}
 p(n_S,n_I) \nonumber \\
 & & \mbox{} \times  K^{S}(c_S,n_S) K^{I}(c_I,n_I) ;
 \label{prob}
\end{eqnarray}
\begin{eqnarray}
  K^{i}(c_i,n_i) &=& \sum_{l=0}^{\min (c_i,n_i)}
  \pmatrix{n_i \cr l\cr} (T_i\eta)^l (1-T_i\eta)^{n_i-l} \nonumber
 \\
  & & \mbox{} \times
   \frac{ D^{c_i-l}}{(c_i-l)!}  \exp(-D) , \hspace{0.8cm} i=S,I .
\end{eqnarray}
This formula holds provided that the number of photons detected
by the camera is much lower than the number of active pixels
\cite{PerinaJr2003}.

\begin{figure}  
 {\raisebox{4 cm}{a)} \hspace{5mm}
 \resizebox{0.6\hsize}{!}{\includegraphics{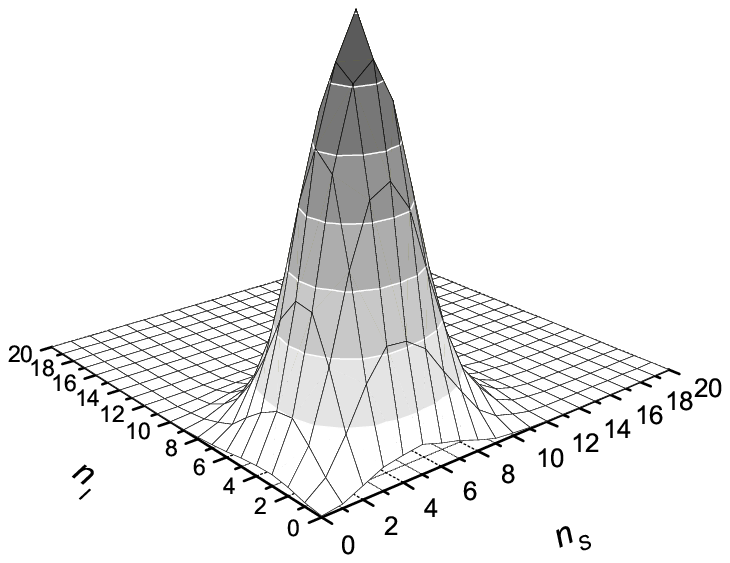}}}

 \mbox{ } \vspace{2mm}

 {\raisebox{4 cm}{b)} \hspace{5mm}
 \resizebox{0.6\hsize}{!}{\includegraphics{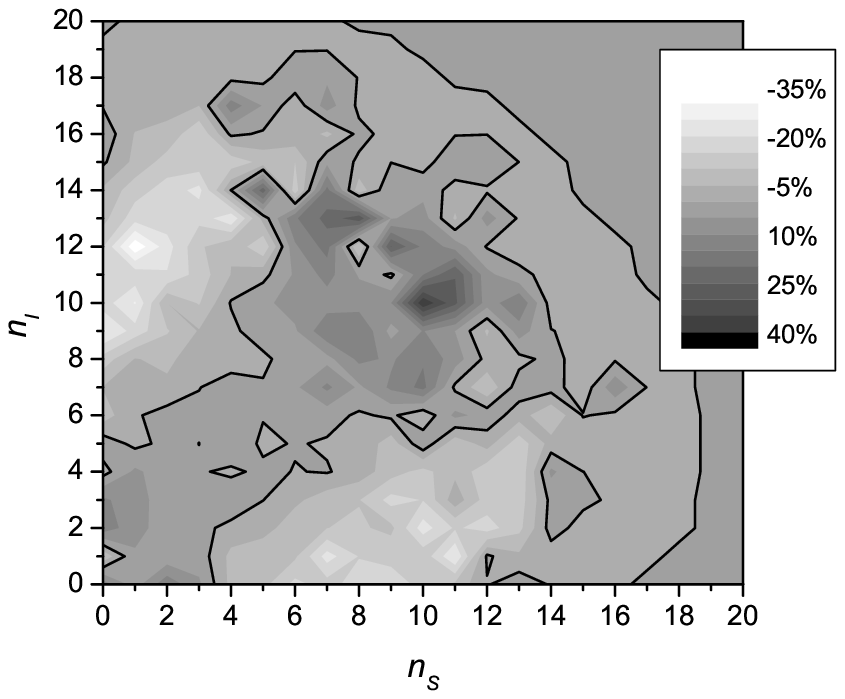}}}
  \caption{(a) Measured joint signal-idler photon-number
  distribution
  $ f $ as a function of the signal
  ($ n_S $) and idler ($ n_I $) photon numbers,
  (b) difference between the measured photon-number distribution
  $ f $ and the distribution given by direct product of two independent
  Poissonian distributions with means equal to those of
  experimental data. Solid line denotes zero contour.}
 \label{pn12}
\end{figure}

An example of the measured photon-number distribution $ f(c_S,c_I)
$ is given in Fig.~\ref{pn12}a. A correlated character of twin
beams is clearly visible in the graph in Fig.~\ref{pn12}b showing
the difference between the measured photon-number distribution $
f(c_S,c_I) $ and that one composed of two independent Poissonian
distributions with mean values equal to those of the measured
signal and idler fields. We can see that elements lying on
diagonal or near diagonal are enhanced whereas those lying far
from diagonal are significantly suppressed. This is a direct
experimental manifestation of the fact that signal and idler
photons are generated in pairs.

Covariance $ C_p $ of the signal and idler photon numbers determined
along the formula \cite{Saleh1978}
\begin{eqnarray}          
  C_p &=& \frac{ \langle \Delta n_S \Delta n_I \rangle }{
  \sqrt{ \langle (\Delta n_S)^2
    \rangle \langle (\Delta n_I)^2 \rangle }} ,
      \label{12} \\
    & & \Delta n_i = n_i - \langle n_i \rangle,
      \vspace{1cm} i=S,I  \nonumber
\end{eqnarray}
equals $0.0435\pm0.008$ for the data shown in Fig.~\ref{pn12}.
Type of photon-number statistics can be judged according to the
value of coefficient $ K $,
\begin{equation}     
 K = \frac{ \langle n^2 \rangle }{\langle n\rangle^2}
  - \frac{1}{\langle n\rangle};
\end{equation}
$ K = 1 $ for Poissonian statistics, $ K = 2 $ characterizes
Gaussian statistics, and $  K < 1 $ for a Fock state.
The experimentally determined marginal signal-field (idler-field)
photon-number distribution has $ K_S $ ($ K_I $) equal to $ 0.997
\pm 0.030 $ ($ 0.994 \pm 0.030 $), i.e. the statististics can be
considered to be Poissonian within the experimental error.
Statistics of photon pairs deviate from the Poissonian ones if the
number of independent modes constituting the field is small
\cite{Riedmatter2003}.

Taking into account the fact that the measured marginal signal and
idler photon-number distributions are Poissonian, the measured
nonzero value of covariance $ C_p $ also
indicates nonclassical properties of detected twin beams. We note
that a classical coherent field (with Poissonian statistics)
cannot have any correlations in photon numbers at two distinct
spatial points.

Pair character of twin beams is clearly revealed in the original
joint signal-idler photon-number distribution $ p(n_S,n_I) $
characterizing the light field as it occurs at the output plane of
the crystal. The method of maximum likelihood estimation has
proven to be extraordinarily suitable in finding the inversion to
the relation in Eq. \ref{prob}. The joint signal-idler
photon-number distribution $ \rho^{(\infty)}(n_S,n_I) $ that
minimizes the Kullback-Leibler divergence from experimental data
can be found using the iterative Expectation-Maximization
algorithm \cite{Dempster1977}:
\begin{eqnarray}    
\rho^{(n+1)}(n_S,n_I) &=& \rho^{(n)}(n_S,n_I)
  \nonumber \\
 & & \hspace{-2.8cm} \times \sum_{i_S,i_I=0}^{\infty}
  \frac{ f^{\infty,\infty}(i_S,i_I)
K^{S,\infty}(i_S,n_S) K^{I,\infty}(i_I,n_I) }{
\sum_{j_S,j_I=0}^{\infty} K^{S,\infty}(i_S,j_S)
K^{I,\infty}(i_I,j_I) \rho^{(n)}(j_S,j_I) } .
 \nonumber \\
 \label{likelihood}
\end{eqnarray}
The symbol $ \rho^{(n)}(n_S,n_I) $ denotes the joint signal-idler
photon-number distribution after the $ n $-th step of iteration,
$ \rho^{(0)}(n_S,n_I) $ is an arbitrary initial photon-number
distribution.

\begin{figure} 
 {\raisebox{4 cm}{a)} \hspace{5mm}
 \resizebox{0.6\hsize}{!}{\includegraphics{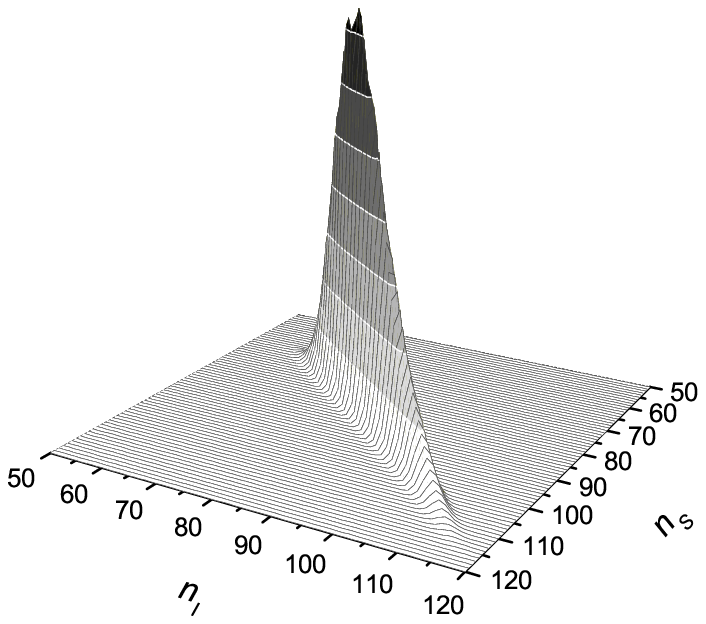}}}

\mbox{} \vspace{2mm}

 {\raisebox{4 cm}{b)} \hspace{5mm}
 \resizebox{0.55\hsize}{!}{\includegraphics{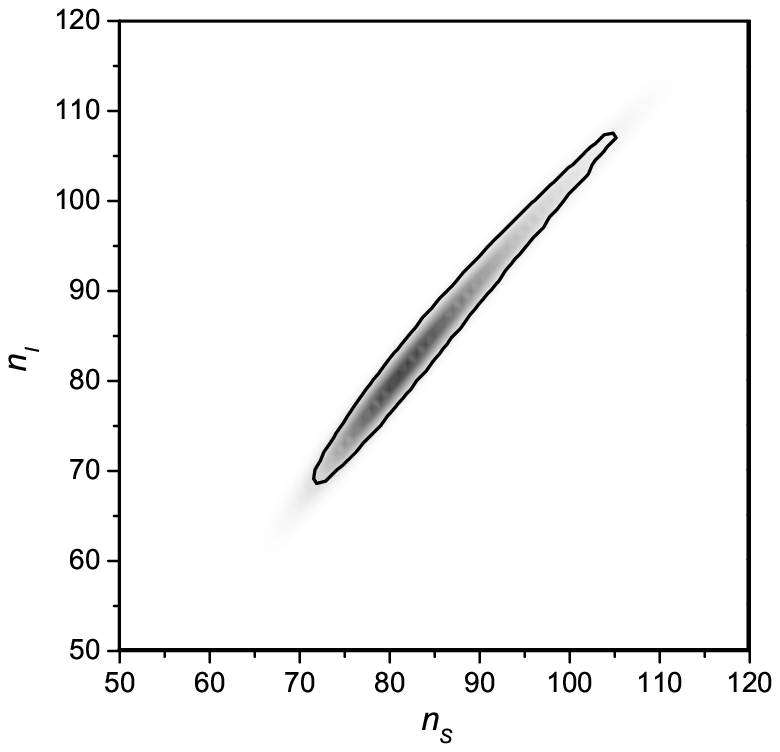}}}
 \vspace{0mm}
  \caption{(a) 3D- and (b) topo-graph of reconstructed joint
  signal-idler photon-number
  distribution $ \rho^{(\infty)} $ as a function of the signal
  ($ n_S $) and idler ($ n_I $) photon numbers; area in graph
  (b) bordered by the solid line contains probabilities violating
   classical inequality in Eq.~\ref{6};
  $ \eta T_S=0.0539 $, $ \eta T_I=0.0415 $, $ D=0.75 $.}
  \label{reconstruction}
\end{figure}

The joint signal-idler photon-number distribution
$\rho^{(\infty)}(n_S,n_I)$ at the output plane of the crystal
determined using Eq.~{\ref{likelihood}} shows strong correlations
between the signal photon numbers and idler photon numbers (see
Fig.~\ref{reconstruction}).
Values of the parameters $ \eta T_S $, $ \eta T_I $, and $ D $
necessary for reconstruction have been determined from an
independent measurement. Covariance of the signal and idler photon
numbers equals $C_p \approx 0.98$ after the reconstruction and
shows the ability of the reconstruction method to reveal the
original pairing of photons. This reconstruction is not perfect
due to the impossibility to describe precisely all noises
occurring in the experiment. Mixing the signal and idler fields
together, the resultant photon-number distribution preferring even
photon-numbers also reflects these correlations, as has been observed
in \cite{Waks2004} using a special photon-number resolving
detector.

Any photon-number distribution $ p(n_S,n_I) $ originating
in a classical field has to fulfil the following inequality
\cite{Hillery1987}:
\begin{equation}    
 p(n_S,n_I) \le \frac{n_S^{n_S}}{n_S!} \exp(-n_S) \, \frac{n_I^{n_I}}{n_I!}
  \exp(-n_I) ;
 \label{6}
\end{equation}
i.e. this inequality represents a criterion of nonclassicality.
The reconstructed probabilities lying inside the bold curve in
Fig.~3b exhibit violation of this inequality. The violation is a
consequence of the generation of photons in pairs (probabilities
tend to be concentrated towards the diagonal where they reach
greater values).

Experimental marginal signal and idler photon-number distributions
derived from the reconstructed joint signal-idler photon-number
distribution are Poissonian ($ K_S = 0.997 $, $ K_I = 1.000 $).
This means that also the probability
distribution of generated photon pairs is Poissonian. Poissonian
statistics reflects the fact that photon pairs can be generated
into many independent modes distinguishable in space and time.
In case of our experimentally obtained twin beams having
typically tens or hundreds of photon pairs, each pair occupies
its own mode with a high probability.

The reconstructed joint signal-idler photon-number distribution
$ \rho^{(\infty)} $ enables to determine the joint distribution
of integrated
intensities $ W_S $ and $ W_I $ of the signal and idler fields
at the output plane of the crystal;
\begin{equation}        
 \hat{W}_l = \int_{-\infty}^{\infty} d\tau \hat{E}^{(-)}_l(\tau)
 \hat{E}^{(+)}_l(\tau) , \hspace{1cm} l=S,I.
  \label{intint}
\end{equation}
Inverting photodetection equation we have for the joint
signal-idler integrated-intensity distribution $ P(W_S,W_I,s,M) $
related to $ s-$ordering of field operators and assuming $ M $
independent modes (\cite{Perina1991}, chap.~4):
\begin{eqnarray}       
 P(W_S,W_I,s,M) &=& \frac{1}{W_S W_I}
 \exp\left[ - \frac{2(W_S + W_I)}{1-s} \right]
 \nonumber \\
 & & \hspace{-3.5cm} \times \left[
 \frac{4 W_S W_I}{(1-s)^2}\right]^M
 \sum_{n_S=0}^{\infty} \sum_{n_I=0}^{\infty}
 \frac{\rho^{(\infty)}(n_S,n_I)n_S! \, n_I!}{[\Gamma(n_S+M)
 \Gamma(n_I+M)]^2}
  \nonumber \\
 & & \hspace{-3.5cm} \times \left( \frac{s+1}{s-1}\right)^{n_S+n_I}
 L^{M-1}_{n_S} \left( \frac{4W_S}{1-s^2} \right)
 L^{M-1}_{n_I} \left( \frac{4W_I}{1-s^2} \right) ;
\end{eqnarray}
$ L^{M-1}_n $ are Laguerre polynomials and $ \Gamma $
means the gamma function. The parameter $ s $ equals -1,
0, and 1 for anti-normal, symmetric, and normal ordering of field
operators, respectively.

\begin{figure}  
 {\raisebox{3.6 cm}{a)} \hspace{5mm}
 \resizebox{0.52\hsize}{!}{\includegraphics{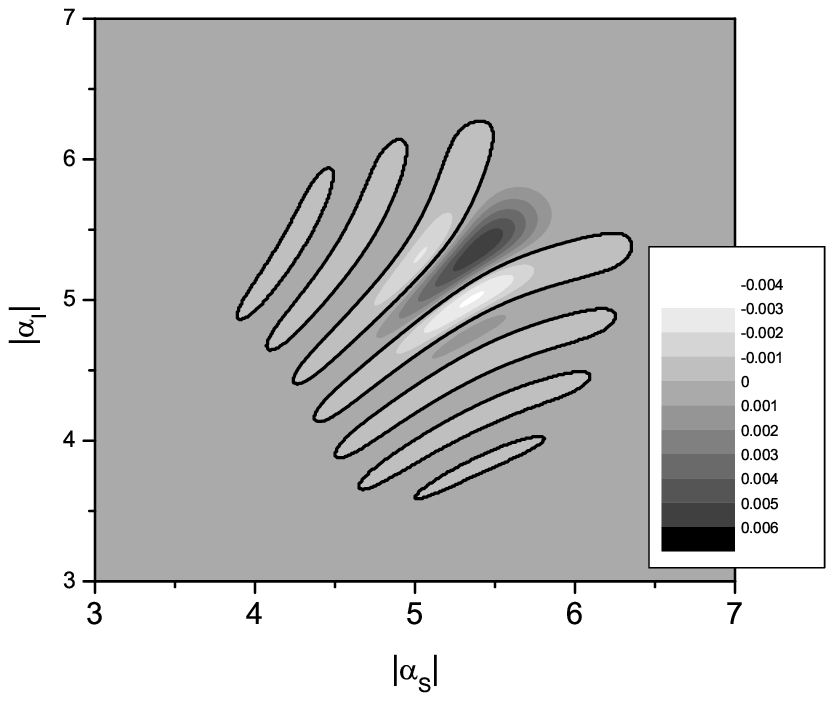}}}

 \mbox{ } \vspace{2mm}

 {\raisebox{3.6 cm}{b)} \hspace{5mm}
 \resizebox{0.52\hsize}{!}{\includegraphics{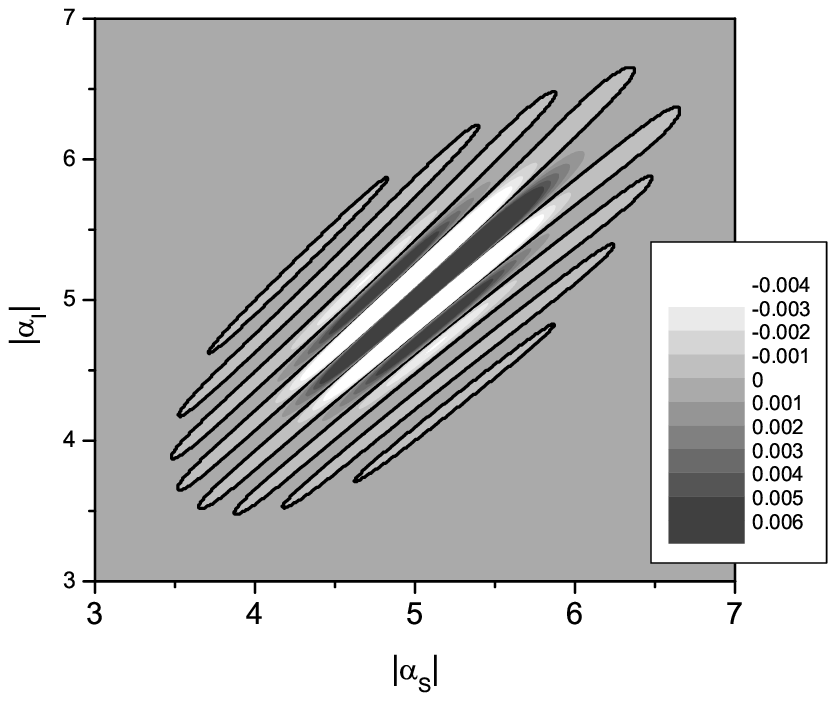}}}
 \caption{Joint signal-idler quasidistribution
  $ \Phi(|\alpha_S|,|\alpha_I|,s=0.3,M=50) $
  determined from (a) experimental data and (b)
  assuming an ideal signal-idler photon-number distribution
  $ p(n_S,n_I) = \delta_{n_S,n_I} \exp(-\mu) \mu^{n_S}/n_S! $,
  $ \mu = 14.5 $.}
 \label{quasidistribution}
\end{figure}

Since both signal and idler fields are phase independent (they are
generated in a multimode spontaneous process), the joint signal-idler
phase-space quasi-distribution $ \Phi(\alpha_S,\alpha_I,s,M) \equiv
\Phi(|\alpha_S|,|\alpha_I|,s,M)$ related to $ s- $ordering can be
simply determined:
\begin{equation}      
 \Phi(|\alpha_S|,|\alpha_I|,s,M) = \frac{1}{\pi^2}
  P(|\alpha_S|^2,|\alpha_I|^2,s,M)
\end{equation}
using integrated-intensity distribution $ P $. A nonclassical
field manifests itself by negative values in quasi-distributions
with $ s \ge s_0 $, where the value of $ s_0 $ is a characteristic
of the state. Quasi-distribution with $ s = 0.3 $ obtained from a
reconstructed photon-number distribution $ \rho^{\infty}(n_S,n_I)
$ and plotted in Fig.~\ref{quasidistribution}a has a typical
structure of correlated states composed of regions with positive
and negative values. For comparison, the same quasi-distribution $
\Phi_{\rm ideal}(|\alpha_S|,|\alpha_I|,s,M) $ belonging to an
ideal (perfectly correlated) Poissonian joint signal-idler
photon-number distribution is shown in
Fig.~\ref{quasidistribution}b. Regions with negative values are
the evidence of nonclassical nature \cite{Lvovsky2001} of twin
beams composed of many photon pairs.

The iCCD camera is also able to reveal spatial correlations of
photon pairs \cite{Jost1998} that can be characterized by
entanglement area of two photons in a photon pair. Spatial
correlations can then be used for further studies of photon-number
statistics. These investigations are in progress.

In conclusion, we have experimentally demonstrated that
twin beams composed of many photon pairs are of nonclassical
origin that is manifested both in joint photon-number distributions
and joint quantum phase-space quasi-distributions.

The authors acknowledge support by the projects Research
Center for Optics (LN00A015) and CEZJ-14/98 of
the Ministry of Education of the Czech Republic.

\end{document}